\documentclass[10pt]{article}

\usepackage{graphics}
\usepackage{epsfig}
\usepackage{latexsym,amssymb}
\usepackage{amsmath}

\begin{document}

\author{ A. Guarino$^{\mathsf{* \blacklozenge}}$, S. Ciliberto$^{\mathsf{*}}$, A. Garcimart%
\'{\i}n$^{\mathsf{\ddagger }}$, \\
M. Zei$^{\mathsf{\bigstar \clubsuit}}$ and R. Scorretti$^{\mathsf{*\heartsuit}}$\\
$^{\mathsf{*}}${\small Ecole Normale Superieure de Lyon, 46 all\'{e}e
d'Italie, 69364 Lyon, France}\\
$^{\mathsf{\ddagger }}${\small Departamento de F\'{\i}sica, Facultad de
Ciencias, Universidad de Navarra,}\\
{\small \ E-31080 Pamplona, Spain.}\\
$^{\mathsf{\clubsuit}}${\small Facolt\`a di Ingegneria, Universit\`a degli
Studi di Firenze,}\\
{\small via S. Marta 4, Firenze, Italia.}\\
}
\title{ Failure time and critical behaviour of fracture precursors in
heterogeneous materials}
\date{}
\maketitle

\begin{abstract}
The acoustic emission of fracture precursors, and the failure time of
samples of heterogeneous materials (wood, fiberglass) are studied as a
function of the load features and geometry. It is shown that in these
materials the failure time is predicted with a good accuracy by a model of
microcrack nucleation proposed by Pomeau. We find that the time interval $%
\delta t$ between events (precursors) and the energy $\varepsilon$ are power law
distributed and that the exponents of these power laws depend on the load
history and on the material. In contrast, the cumulated acoustic energy $E$
presents a critical divergency near the breaking time $\tau $ which is $%
E\sim \left( \frac{\tau -t}\tau \right) ^{-\gamma }$. The positive exponent $%
\gamma $ is independent, within error bars, on all the
experimental parameters.
\end{abstract}

\section{Introduction}

Heterogeneous materials are widely studied not only for their
large utility in applications but also because they could give
more insight to our understanding of the role of macroscopic
disorder on material properties. The statistical analysis of the
failure of these materials is an actual and fundamental problem
which has received a lot of attention  both theoretically
\cite{hermann,sornette,sha,libro,atki,naimark,zei} and
experimentally\cite{Anifrani,prl,articolo,stead,maes}. When an
heterogeneous material is stretched its evolution toward breaking
is characterized by the appearance of microcracks before their
final break-up. Each microcrack produces an elastic wave which is
detectable by a piezoelectric microphone. The microcraks
constitute the so called precursors of fracture. It is very well
known that these materials subjected to a constant load may break
after a certain time, which is a function of the applied load.
Many models have been proposed to predict this failure time, but
the physical mechanisms remain
unclear\cite{sha,libro,Zhurkov,Lawn}. Very recently it has been
proposed \cite{golubovic,pom} a model, which
explains quite well the failure time of microcrystals \cite{pauch} and gels %
\cite{bonn} submitted to a constant stress load. This model is
based on the idea that a nucleation process of a microcrack has to
take place inside the materials, in order to form the macroscopic
crack. This nucleation process is controlled by an activation law,
as the coalescence of
a phase into another in a liquid-solid transition. Based on this prediction %
\cite{pom}, L. Pauchard et al. \cite{pauch} found that if a
constant load is applied to a bidimensional
microcrystal\cite{commentaccio}, it breaks after a time $\tau $
given by the equation $\tau =\tau _{o}e^{P_{o}^{2}/P^{2}}$, where
$P$ is the
applied pressure, and $\tau _{o}$ and $P_{o}$ are constants. Bonn et al. %
\cite{bonn} found a similar law for gels. Pomeau predicted that for
three-dimensional microscopic systems the life-time should be:
\begin{equation}
\tau =\tau _{o}\exp \left( \frac{P_{o}}{P}\right) ^{4}  \label{const}
\end{equation}
where $\tau _{o}$ is a characteristic time and $P_{o}$ a
characteristic pressure, which mainly depend on the material
characteristics, the experimental geometry and temperature.
The idea, that the life time of a material might be due to a
 thermally activated process,has been proposed long time ago by Mogi \cite{mogi} and Zhurkov
 \cite{Zhurkov}. They got a different  expression for $\tau$
 \begin{equation}
\tau =\tau _{o}\exp \left(- \frac{P}{P_{o}}\right) \label{eqmogi}
\end{equation}
This equation   was accurately checked  in many homogeneous
materials and it shows  a good agreement with experimental data
\cite{Zhurkov}. However it has to be stressed that eq.\ref{const}
and \ref{eqmogi} are certainly first order approximation because
they neglect the tensorial nature of crack perturbations and their
long range  interactions \cite{naimark}. These ideas are quite
interesting  and it is important to check experimentally
 whether they can be applied in heterogeneous materials, such as fiber glass and wood
pannels\cite{nature}. In two recent papers \cite{prl, articolo},
we have shown that in these materials the microcracks, preceding
the main crack form something like a coalescence around the final
path of the main crack. The purpose of this paper is to
investigate more deeply the behaviour of these materials, and
specifically the statistical properties of fracture precursors and
their relationships with the failure time. The paper is organized
as follows: in section 2 we describe the different experimental
settings. In section 3 we study the time failure for the samples
submitted to a constant load (3.1) and its statistical properties
(3.2). Then we generalize to a time dependent load. In section 4
the statistical behaviour of the fracture precursors is studied as
a function of the load features and the geometry. Discussion and
conclusions follow in section 5.

\section{\protect\bigskip Experimental setups}

In order to verify the dependence of the results on the geometry
and on the fracture's mode, we used three different experimental
setups. We performed mode-I fracture experiments both with a
classical tensile machine (TM) and a high pressure chambers (HPC).
In the case of the TM the stress distribution is very simple but,
due to moving mechanical parts, we have to deal with a large
acoustic noise. In order to avoid noise and to detect reliably
microcracks with a weak sound emission, we have designed a set-up
in which there are no moving parts except the sample itself, the
HPC. Here, the stress distribution is very complicated but
numerical calculations\cite{tesi-guarino} show that the experience
can be thought of as a Mode-I test with circular symmetry.
Finally, we used a flexion-machine to perform mode-I tests.

\subsection{Samples}

Several materials have been used. Most of the runs have been
carried out on two fibrous composite materials: chipboard wood
panel, which is made of small wood fibers randomly oriented, and
different fiberglass panels made of a fiber fabric and an epoxy
resin. The Young modulus of the samples is $1.8\ 10^{8}\ N/m^{2}$
and $10^{10}\ N/m^{2}$ for the wood panels and the fiberglass
respectively. The Poisson's modulus is $\nu = 0.35$ for both materials.
The longitudinal sound velocity is 1900$\ $ $m/s$
for wood panels and 2200$\ $ $m/s$ for fiberglass. The choice of
the materials was determined by their features: they consist of
small fibers, randomly oriented, and they are elastic and
heterogeneous. The geometry of the samples, which depends on the
experimental set up used to test it, is described in the following
sections.

\subsection{High pressure chambers (HPC)}

A circular wood or fiberglass sample having a diameter of $22$
$cm$ and a thickness between $1$ and $5$ $mm$ is placed between
two chambers between which a pressure difference $P=P_{2}-P_{1}$
is imposed(see fig. \ref{fig:macchine}b). If the deformation of
the plate at the center is bigger than its thickness, which is the
case here, the load is mainly radial \cite{Landau,Timoshenko}.
Therefore, the experience can be thought of as a Mode-I test with
circular symmetry. The pressure difference $P$ supported by the
sample is slowly increased and it is monitored by a differential
transducer. This measure has a stability of $0.002$ $atm$. The
fracture pressure for the different tested materials ranges from
$0.7$ to $2$ $atm$. We regulate $P$ by means of a feedback loop
and an electronically controlled valve which connects one of the
two chambers to a pressurized air reservoir. The time taken to
correct pressure variations (about 0.1 second) is smaller than the
characteristic time of the strain rate. An inductive displacement
sensor (Linear Differential Variable Transducer 500HR from PM
Instrumentation) gives the deformation at the center of the plate
with a precision of about 10 microns (the deformation just before
fracture is of the order of one centimeter, depending on the
material). The apparatus is placed inside a copper box covered
with a thick foam layer to avoid both electrical and acoustic
noise. Four wide-band piezoelectric microphones (Valpey-Fisher
Pinducer VP-1093) are placed on the side of the sample (see fig.
\ref{fig:macchine}a,b). The signal is amplified, low-pass filtered
at $70$ $kHz$ , and sent to a digitizing oscilloscope and to an
electronic device which measures the acoustic energy detected by
the microphones. The signal captured by the oscilloscope is sent
to a computer where a program automatically detects the arrival
time of the acoustic emissions (AE) at each microphone. If the
signal is detected by more than two microphones, a calculation
yields the position of the source inside the sample. A fraction of
the detected events are rejected, either as a result of a large
uncertainty of the location, or because they are regarded as
noise. The mean standard error for the calculated positions is
about $6$ $mm$, which results mainly from the uncertainty of the
arrival time. The electronic device that measures the energy
performs the square of the AE amplitude and then integrates it
over a time window of $30$ $ms$, which is the maximum duration of
one acoustic event. The output signal is proportional to the
energy of the events, and its value is sent to the computer. The
dynamic range for the energy measurement is four decades, and the
device is adjusted in such a way that only the strong sound
emitted by the final crack saturates it. The global results of the
measurements are the following: a list of the positions of
microcracks, the strain of the samples and the energy released as
a function of the control parameter $P$. Further details of the
setup are described elsewhere\cite{articolo, prl, nature,
tesi-guarino}.

\subsection{Tensile Machine (TM)}

The experimental apparatus consists of a tensile machine (see fig.
\ref{fig:macchine}c) which can apply a maximum force of about
$23000$ $N$. During the load we measure the applied force $F$, the
strain, the $AE$ produced by microcracks and the time at which the
event was detected. The data acquisition set-up is the same used
with the HPC. The samples have a rectangular shape of size $l = 30$ $cm$, $w = 20$ $cm$ and thickness of $2$ $mm$.
More details of the experimental setup can be found in \cite{boudet}.

\subsection{Bending-machine (BM)}

The apparatus is a three-points flexion machine,
fig.\ref{fig:flexmach}. The rectangular sample lies horizontally
with fixed edges and vertical load is imposed in its center. The
size of the samples is $l = 8$ to $22$ $cm$, $w = 1$ to $2$ $cm$,
and the thickness is $0.2$ $cm$. With this setup we only used fiberglass samples. We
can load the sample up to $65Kg$ (that is the machine critical
load) by minimum steps of $100$ $g$. An inductive displacement
sensor, similar to the one used in the HPC, has been used to
measure the displacement of the center of the sample $f$. The
sensor is connected to a computer that samples the signal at
$1Hz$. The failure time $\tau $ is obtained by the analysis of the signal $%
f(t)$; The uncertainty of $\tau $ is then $0.5$ $s$. No acoustic
emissions are measured with this apparatus. More details are
described in \cite{tesimaria}.

\section{The failure time}

The aim of this section is to study the lifetime $\tau$ of heterogeneous
materials and to check whether eq.(1) could be useful in order to
predict it. We first investigate $\tau $ when samples are submitted
to a constant load using the $HPC$ and the $BM$. Then we consider
the case of a time dependent load, and we try to generalize
eq.(1).

\subsection{Constant load}

We first impose a constant strain to our samples (using all the
apparatus), as it has been made for crystals\cite{pauch}. As
strain is fixed, every microcrack leads to a pressure decrease: in
fact each microcrack weakens the material, so that a lower
pressure is needed to keep the strain constant. It follows that if
the imposed strain is small enough\footnote{it's clear that if the
imposed strain has to be smaller than a critical value, at which
the sample breaks instantaneously.}, the system reaches a
stationary state, where the pressure remains constant and no more
microcracks are detected\cite{tesi-guarino} (see fig. \ref{fig:ultima}b).
One sample was submitted to a large deformation (i.e. close to the critical
value) and it did not break after three days. Therefore, at
imposed strain, the effect observed in microcrystals is not valid
for heterogeneous materials.

On the other hand, if a constant stress is imposed to the system,
no matter which apparatus we use, it will break after a certain
time which depends on the value of the applied load. This can be
done by imposing a constant pressure with the HPC (see fig.
\ref{fig:ultima}a) or a constant force with the TM and BM. The
reason for this is that after every single microcrack the same
load must be endured by the weakened sample, so that it becomes
more and more unstable. Using either the TM or HPC, we have
submitted several samples to different constant loads $P$ and we
have measured the life-time $\tau $. The values obtained are well
fitted by eq.(\ref{const}), that is the exponential function
predicted by Pomeau. On the other hand, the life-time expression
$\tau =ae^{-bP}$ proposed by Mogi\cite{mogi} does not conform to
our data\cite{commento}. The same law has been found
experimentally by Zhurkov\cite{Zhurkov}. However it's worth noting
that his work deals mainly with homogeneous, visco-elastic and
plastic materials, whereas the materials we used are heterogeneous
and elastic: this could explains why he found a different
dependence of $\tau$ on the imposed load. In fig. \ref{fig:mogi}
$\tau $ is plotted versus $\frac{1}{P^{4}}$ in a semilog scale and
a straight line is obtained. Fig.\ref{fig:mogi} corresponds to the
case of wood samples broken in the HPC, while in fig.
\ref{fig:mariap4a} we show the points for fiberglass samples
broken in the BM. Each point corresponds to the mean value
obtained with $20$ samples. Even if the load is very small the
sample will eventually break, although the life-time can be
extremely long. For example, using eq.(\ref{const}) and the best
fit parameters of fig. \ref{fig:mogi}a, one estimates $\tau \simeq
5000$ s at $P=0.43$ atm. Halving the imposed pressure causes $\tau
$ to become extremely large : $\tau =4.4\cdot 10^{37}$ years at
$P=0.21$ atm).

The value of $\tau_o$ seems to depend on the geometry and the
material but not, within the error bar, on the sample size. In
fact, for the circular samples broken with the HPC, we find $\tau
_{o}=50.5\pm 0.2$ $s$ for wood and $\tau _{o}=44.6\pm 0.2$ $s$ for
fiberglass. While in the the BM we find $\tau_o=2.5\pm 0.3$ $s$
and $\tau_o=2.7\pm 0.3$ $s$ for the samples with $W=1  cm$ and
$W=2 cm$ respectively.

The value of $P_o$ depends both on the geometry, the size and the
material of the sample, and indeed we find $P_o=2.91 \ atm$ and
$P_o=0.63 \ atm$ for fiberglass and wood broken in the $HPC$. For
the fiber glass samples broken with the BM the values of $P_o$ are
$P_{o}=71.1 \ kg$ for $W=2 \ cm$ and $P_o=35 \ kg$ for $W=1 \ cm$

\subsection{Statistic of the failure times}

It's interesting to study the statistical distribution $N(\tau)$
of the failure times. This information can give more insight on
the physical phenomena.

The main hypothesis of Pomeau is that the failure of a sample is
due to the thermal nucleation of one defect. Thus one expects that
the failure time $\tau$ follows a Poisson's distribution. This has
been experimentally observed in gels\cite{bonn}. Conversely, in
all the measures made with the HPC and the TM the failure time
follows a normal distribution. This is also the case for
crystals\cite{pauch}.

This difference can be explained by the fact that in our samples
the failure is due to the nucleation and coalescence of a large
number of defects, each of one is thermally activated and would
eventually follows a Poisson's law, if it were isolated. Numerical
simulations and analytical calculations  seem to confirm this
idea\cite{roux,noi-physicad,noi-epl2}.

In the case of the experiments performed with the BM, the failure
time distribution $N(\tau)$ seems not to follow a Poisson's law
nor a normal distribution,fig. \ref{fig:mariahist}a. In this case
we observed that the cumulative distribution
$$Q(\tau) =
\frac{\int_0^\tau N(t)dt}{\int_0^\infty N(t)dt}$$
 is best fitted
by the sum of two exponentials, fig.\ref{fig:mariacumhist}b. We
believe that this is due to the fact that the two components of
these samples (the resine and the texture of glass fibers) give
rise to different characteristic times. We think that this
"separation effect" is observed only with the BM because of the
small size of the samples.

\subsection{Time dependent load}

In order to find a law that holds for a time dependent imposed
stress, we intend to generalize the eq.(\ref{const}). If the
pressure $P$ changes with time, it is reasonable to consider the
entire history of the load. Therefore we consider that
\begin{equation*}
{\frac 1{\tau _o}}\exp \left[-\left(\frac{P_o}P\right)^4\right]
\end{equation*}
is the density of damage per unit time, where $\tau_o$ and $P_o$
 are fitting parameters obtained in the constant load case. The certitude of breaking is
obtained after a time $\tau $ such that:
\begin{equation}
\int_0^\tau \frac 1{\tau _o}\exp\left[-\left(\frac{P_o}P\right)^4\right]dt=1
\label{integ}
\end{equation}
where $\tau _o$ and $P_o$ have the previously determined value.
Notice that this equation is equivalent to eq.(\ref{const}) when a
constant pressure is applied.

To check this idea, we have applied the load to the sample (using
the HPC) following different schemes. We have first applied
successive pressure plateaux in order to check whether memory
effects exist. In fig. \ref{fig:tempi_di_rottura}a the pressure
applied to the sample is shown as a function of time. A constant
load has been applied during a certain time $\tau _1$, then the
load is suppressed and the same constant load is applied again for a time
interval $\tau _2$. The sample breaks after a
loading a time $\tau _1+\tau _2$ which is equal to the time needed
if the same load had been applied continuously without the absence
of load during a certain interval. Therefore a memory of the load
history exists. The life-time formula (eq. \ref{integ}) is also
valid if different constant loads are applied successively (fig.
\ref{fig:tempi_di_rottura}b). This concept can explain the
violation of the Kaiser effect in these materials\cite{articolo}.

If the load is not constant, the life-times resulting from the
proposed integral equation are still in good agreement with
experimental data. A load linearly increasing at different rates
$A_p$ has been applied to different samples. The measured breaking
times are plotted in fig. \ref{fig:tempi_legno_fibra} along with a
curve showing the values computed from eq.\ref{integ}. Even if a
quasi-static load is applied erratically (fig.
\ref{fig:tempi_di_rottura}c), the calculated life-time agrees with
the measured one. These experiments show that eq. \ref{integ}
describes well the life-time of the samples submitted to a time
dependent pressure.

\subsection{The dependence of $\tau$ on the temperature}

The question is to understand why eq.(\ref{const}) and
(\ref{integ}) works so well for a three dimensional heterogeneous
material. Indeed, in the Pomeau formulation \cite{pom}
\begin{equation}
P_o=G \left( \frac{\eta ^3Y^2}{KT}\right) ^{1/4}  \label{PO}
\end{equation}
where $Y$ is the Young modulus, $T$ the temperature, $K$ the
Boltzmann constant and $\eta $ the surface energy of the material
under study. $G$ is a geometrical factor which may depend on the
experimental geometry, on defect shape and density.

In our experiment with HPC, we found $P_o=0.62$ $atm$ for wood, which has Y=$%
1.8\cdot10^8$ $N/m^2$, and $P_o=2.91$ $atm$ for fiberglass, which
has $Y = 10^{10}$ $N/m^2$. Thus the ratio between the values of
$P_o$ found for the two materials is closed to the ratio of the
square root of their Young modula.

In contrast temperature does not seem to have a strong influence on $\tau $.
In fact we changed temperature, from $300K$ to $380K$ which is a temperature
range where the other parameters, $Y$ and $\eta $, do not change too much.
For this temperature jump one would expect a change in $\tau $ of of about $%
50\%$ for the smallest pressure and of about $100\%$ for the
largest pressure. Looking at fig. \ref{fig:tempi_legno_fibra} we
do not notice any change of $\tau $ within experimental errors
which are about $10\%$. In order to maintain the change of $\tau $
within $10\%$ for a temperature jump of $80K$ one has to assume
that the effective temperature of the system is about $3000K$.
Notice that this claim is independent on the exact value of the
other parameters and G.

These observations seem to indicate that the nucleation process of
microcracks is activated by a noise much larger than the thermal
one. Such a large noise can be probably produced by the internal
random distribution of the defects in the heterogeneous materials
that we used in our experiments. This internal random distribution
of material defects evolves in time because of the appearance of
new microcracks and the deformation of the sample. Therefore this
internal and time dependent disorder of the material could
actually be the mechanism that activates the microcrack
coalescence and play the role of a very high temperature.
Numerical simulations and analytical calculations, which we
performed in fuse networks,  confirm this hypothesis\cite{roux,
noi-physicad, noi-epl2}. Similar conclusions on the  role of
disorder in  the crack activation processes have been reached by
other authors\cite{Arndt,naimark}.  The  experimental test of
these models is a very important point which merits to be deeply
explored in the future.

\section{Statistical behavior of fracture precursors}

When a constant pressure is applied to the sample using the HPC,
the acoustic emissions of the material are measured as a function
of time. We find that the cumulative acoustic energy $E$ diverges
as a function of the reduced time $\frac{\tau-t}{\tau}$,
specifically $E \propto (\frac{\tau- t}{\tau})^\gamma$ with
$\gamma=0.27$ (see Fig. \ref{fig:critic2}). Notably, the exponent
$\gamma$, found in this experiment with a constant applied
pressure, is the same of the one corresponding to the case of
constant stress rate \cite{prl}. Indeed it has been shown
\cite{prl,articolo} that if a quasi-static constant pressure rate
is imposed, that is $P=A_pt$, the sample breaks at a critical
pressure $P_c$ and $E$ realesed by the final crack precursors
(microcracks) scales with the reduced pressure or time (time and
pressure are proportional) in the following way:
\begin{equation}  \label{Energy}
E \propto \left(\frac{P_c-P}{P_c}\right)^\gamma = \left(\frac{\tau-t}{\tau}%
\right)^\gamma
\end{equation}
where $\tau=P_c/A_p$ in this case. Thus it seems that the real
control parameter of the failure process is time, regardless of
the fact that either a constant pressure rate or a constant
pressure is applied. In the case of constant load rate ($P=A_pt$
or $u=Bt$) the system has not a characteristic scale of energy or
time: the histogram $N(\varepsilon )$ of the released energy and
the histogram $N(\delta t)$ of the elapsed time $\delta t$ between
two consecutive events reveal power laws, i.e.
$N(\varepsilon)\thicksim \varepsilon ^{-\alpha }$ and $N(\delta
t)\thicksim\delta t^{-\beta }$. The exponents $\alpha $, $\beta $
and $\gamma$ do not depend on the load rate $A_p$ or
$B$\cite{prl,articolo}. Power laws for similar magnitudes are
found experimentally on cellular glass\cite{maes}, and numerically
in a related process, the dielectic breakdown\cite{Guinea1}. The
value of the exponents are not too different. We are interested in
studying the exponents in different geometries and when a constant
(creep test), cyclic or erratic load are imposed. To check the
dependence of $\alpha $, $\beta $ and $\gamma $ on the geometry we
used the TM. The force applied to the sample is slowly and
constantly increased till the sample fails. During the loading we
measure the applied force $F$, the strain, the AE produced by
microcracks and the time at which the event was detected.

\subsection{The dependence of $\protect\alpha$ and $\protect\beta$ on the
load features}

In the experiments performed with the HPC, power laws are obtained
for the distributions of $\epsilon $ and of $\delta t$. As an
example of two typical distributions obtained at constant imposed
pressure, we plot in fig \ref{fig:hist}a) and \ref{fig:hist}b)
$N(\delta t)$ and $N(\delta \epsilon)$ respectively. The exponents
of these power laws ($\alpha _c$ for energies and $\beta _c$ for
times) depend on $P$. In fig. \ref{fig:hist}c, $\alpha _c$ and
$\beta _c$ are plotted versus $P$. Note that both exponents grow
with pressure. We observe that the rate of emissions increases
with pressure, so that the
weight of big values of $\delta t$ decreases. This explains the fact that $%
\beta _c$ grows with pressure. We have compared the histograms of energy $%
\epsilon $ for several pressures, and we noticed that the number of
high-energy emissions is almost the same, while the number of low-energy
emissions increase with pressure, so that the exponent $\alpha _c$ increases
as well. Moreover, as the pressure increases, the exponents $\alpha _c$ and $%
\beta _c$ attain the values $\alpha =1.9\pm 0.1$ and $\beta
=1.51\pm 0.05$ obtained in the case of a constant loading
rate\cite{articolo}. We imposed to the sample a erratic and an
cyclic load, which are plotted as a function of time in figure
\ref{fig:press}a and \ref{fig:press}b respectively. Power laws are
obtained for the distributions of $\epsilon $ and for $\delta t$.
The exponents of these power laws do not depend on the load
behavior; their value is the same of that at constant loading
rate. These and previous results \cite{prl,articolo}, allows us to
state that if $\frac{dP}{dt}\neq 0$, the histograms of the
released energy $\varepsilon$ and of the time intervals $\delta t$
do not depend on the load history. The fact that $\alpha $ and
$\beta $ do not depend on $\frac{dP}{dt}$ seems to be in contrast
with the fact that $\alpha _c$ and $\beta _c$ depend on $P$. This
result can be interpreted by
considering that the microcracks formation process is not the same when $%
\frac{dP}{dt}=0$ and $\frac{dP}{dt}\neq 0$. In the former case, imposed
constant $P$, the mechanism of microcrack nucleation is the dominant one and
the nucleation time depends on pressure. In the other case, $\frac{dP}{dt}%
\neq 0$, the dominant mechanism is not the nucleation but the fact that,
when pressure increases as a function of time, several parts of the sample
may have to support a pressure larger than the local critical stress to
break bonds. The fact that at high constant pressure $\alpha _c$ and $\beta
_c$ recover the value $\alpha _c$ and $\beta _c$ has a simple explanation.
Indeed, in order to reach a very high pressure $P_h$, $\frac{dP}{dt}$ is
different from zero for a time interval which is comparable or even larger
than the time interval spent at constant pressure $P_h$. Thus at high
constant pressure the system is close to the case $\frac{dP}{dt}\neq 0$.

\subsection{The dependence of $\protect\gamma$ on the load features}

The measures performed with the HPC imposing an erratic and a
cyclic pressure, plotted respectively in fig. \ref{fig:press}a and
\ref{fig:press}b, allow us to check the dependence of $\gamma $ on
the history of the sample, i.e. on the behavior of the imposed
pressure. The cumulated energy $E$ for the erratic and the cyclic
pressure, shown in fig \ref{fig:press}a and \ref{fig:press}b as a
function of $t$, is plotted in log-log scale as a function of the
reduced parameter $\frac{\tau -t}\tau $ in fig \ref{fig:critic}a
and \ref{fig:critic}b respectively. We observe that, in spite of
the fluctuations due to the strong oscillations of the applied
pressure, near the failure the energy $E$, as a function of $
\frac{\tau -t}\tau $, is fitted by a power law with $\gamma\simeq
0.27 \pm 0.02$. In fig.\ref{fig:critic}c (reproduced in fig.\ref{fig:critic2} for the sake of clearness), $E$ measured when a constant pressure is applied to the sample is plotted as a
function of $\frac{\tau -t}\tau $. A power law is found in this case too \cite{nature}. The exponent $\gamma $ is, within error bars, the same in the three cases. Hence it seems not to depend neither on the applied pressure history nor on the material\cite{prl,articolo,nature}.\newline

Further, experiments made with the TM show that $\gamma$ is
independent on the geometry. In fact we observe that the behavior
of the energy near the fracture as a function of
$\left(\frac{\tau-t}\tau \right)$ is still a power law of exponent
$\gamma\simeq 0.27 $, as shown in figure \ref{fig:critic}d.

\section{Discussion and conclusions}

We presented the results of experiments regarding fracture of
heterogeneous materials (chipboard and fiberglass). When these materials are
submitted to a strain, one observes acoustic emissions
(precursors) before the samples fails. We have measured the
acoustic emissions and the lifetime $\tau$ of the samples.

We have shown that eq.(1) proposed by Pomeau predicts well the
functional dependence of $\tau$ on the applied load. However the
original Pomeau's theory is unable to explain some features that
are observed in our experiments, in microcrystals\cite{pauch} and
gels\cite{bonn}. In fact eq.(1) is based on the idea that the
fracture is due to the nucleation of {\it one} preexisting defect,
which is thermally activated. We have shown that in our
experiments the fracture is due to the nucleation and coalescence
of a large number of defects, as confirmed by the presence of
acoustic emissions and the shape of the statistical distribution
of lifetimes. Moreover, we found that the lifetimes are not
affected, within the experimental errors (20 \%), by the
temperature. We have calculated that in order to estimate the
measured lifetimes, the temperature $T$ to insert in
eq.\ref{const} and in eq.\ref{PO} should be about $3000$ $K$.
Similar results have been found in 2D crystals\cite{commentaccio}
and gels, where the temperature $T$ to insert in eq.(3) should be
about $1000K<T<2500K$ and $T_{eff}>10^{10}K$ respectively. As for
wood and fiberglass, also in gels the lifetime $\tau $ of the
sample is not influenced, in the limit of experimental errors, by
a variation of the temperature $T$ from $20$ to $90$ ${{}^{\circ
}}C$. In contrast, experiments on 2D-crystals \cite{pauch} show
that $\tau $ depends on $T$.

%To explain these points we propose to modify Pomeau's model by
%saying that the fracture is due to the nucleation and coalescence
%of a certain number of defects, each of one is thermally
%activated. In such a model, the lifetime $\tau $ of the sample
%depends on the heterogeneity of the material and the disorder in
%some way enhances thermal fluctuations so that the nucleation time
%of defects becomes of the order of the measured ones.

To explain these points we propose to modify in a statistical
model the one of Pomeau. In this revised model the failure of the
sample is due to the nucleation and coalescence of a certain
number of defects, each of which is thermally activated. The
parameters $Y$, $\eta$ and $T$ of eq.(3), become average
parameters, which keep into account that the interaction between
defects have a tensor nature and are long range \cite{naimark}. To
explain the fact that the thermal temperature has a minor
influence on the lifetime, we suppose that the strong
time-dependent fluctuations of the internal forces induced by the
heterogeneity (defects, microcraks ...) can be considered as a
sort of noise. Thus $T$ depends on the thermal temperature but
mainly on the disorder in the medium. In this way the
heterogeneity of the material enhances thermal fluctuations so
that the nucleation time of defects becomes of the order of the
measured ones. Recently Arndt {\it et al.} reached a similar
conclusion\cite{Arndt} using a different approach, which is
coherent with the generation of a time dependent distribution of
the  microcrack ensemble \cite{naimark}. Numerical simulation that
we have performed on a very simple model\cite{noi-physicad,
noi-epl2} are in agreement with these results. This model allows
us to generalize eq.(1) to the case of a time dependent load, and
to explain the violation of the Kaiser effect in these materials.

We also studied the statistical properties of the fracture precursors.
 We have found that the histograms of the energy and of time between
 two consecutive AE follow power laws of exponents
$\beta$ and $\alpha$ respectively.
In proximity of the fracture the cumulated energy follows a power law,
 typical of phase transitions. Notably, the critical exponent $\gamma$ seems
 to be independent on the geometry and the applied load. Indeed if AE
 is considered as a susceptibility it is not easy to put
 together the observed critical divergency with a nucleation process.
 Probably the standard phase transition description can be only partially
  applied to failure because of the intrinsic irreversibility of the crack formation.

\section{ Acknowledgements}

We thank   P. Metz, M. Moulin and F. Vittoz for very useful
technical assistance.

\medskip Correspondence should be addressed to S. Ciliberto (e-mail:
sergio.ciliberto@ens-lyon.fr)
\newline{\noindent Actual address:}
\newline {\small
\textsc{$\heartsuit$}  Ecole Centrale de Lyon, 36 avenue Guy de
Collongue, 69131 Ecully, France
\newline {\small
\textsc{$\bigstar$} SPCSI, CEA-Saclay, 91191 Gif-sur-Yvette Cedex,
France}
\newline {\small
\textsc{$\blacklozenge$} Universit\'e  de la Polynesie Francaise,
Jeune Equipe Terre Ocean, BP 6570, FAA'A- Tahiti, French Polynesia

\newpage

\begin{figure}[tbp]
\centerline{\epsfysize=0.9\linewidth \epsffile{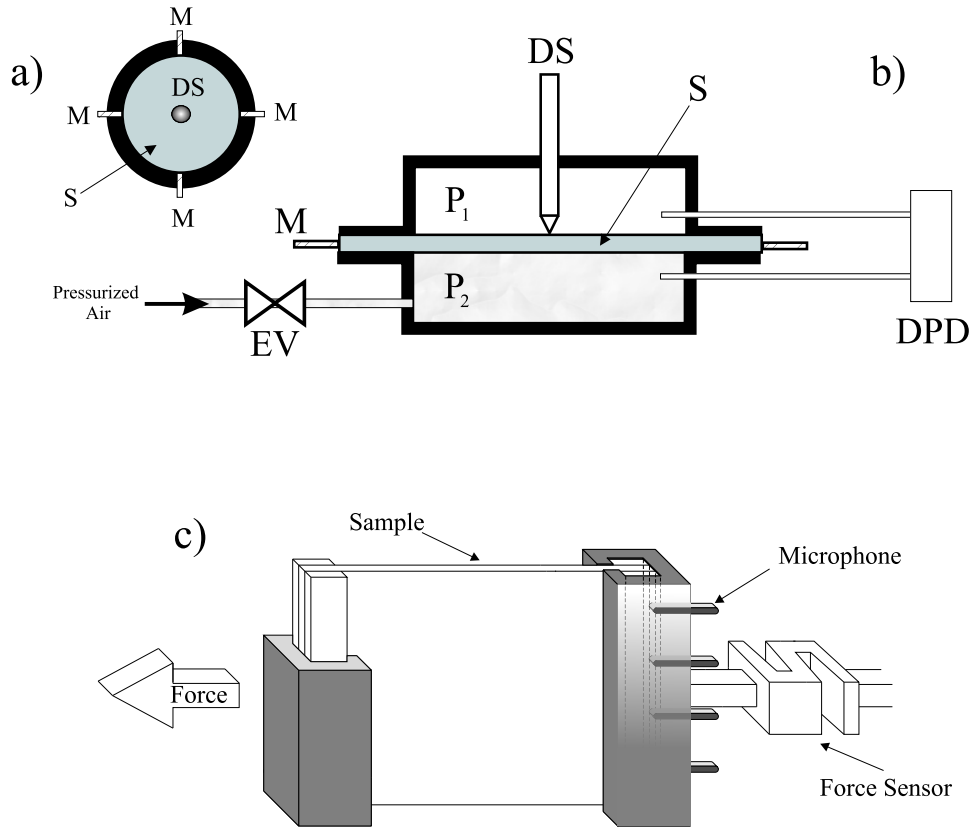}}
\caption{\textbf{a, b) } Sketch of the high pressure chamber (HPC)
apparatus. S is the sample, DS is the inductive displacement
sensor (which has a sensitivity of the order of 1 $\protect\mu
$m). M are the four wide-band piezoelectric microphones.
P=P$_1$-P$_2$ is the pressure supported by the sample. P is
measured by a differential pressure sensor DPD( sensitivity 0.002
atm). EV is the electronic valve
which controls P via the feedback control system Ctrl .
\textbf{c)} Sketch of the tensile
machine. An uniaxial force, which is measured by a piezoresistive
sensor, is applied to the sample by a stepping motor. Four
wide-band piezoelectric microphones measure the acoustic emissions
emitted by the sample. Experiments have been done using
rectangular (20 x 29 cm) wood samples of 4 mm thickness. The whole
apparatus is surrounded by a Faraday screen.} \label{fig:exp}
\label{fig:macchine}
\end{figure}

\begin{figure}[tbp]
\centerline{\epsfysize=0.9\linewidth \epsffile{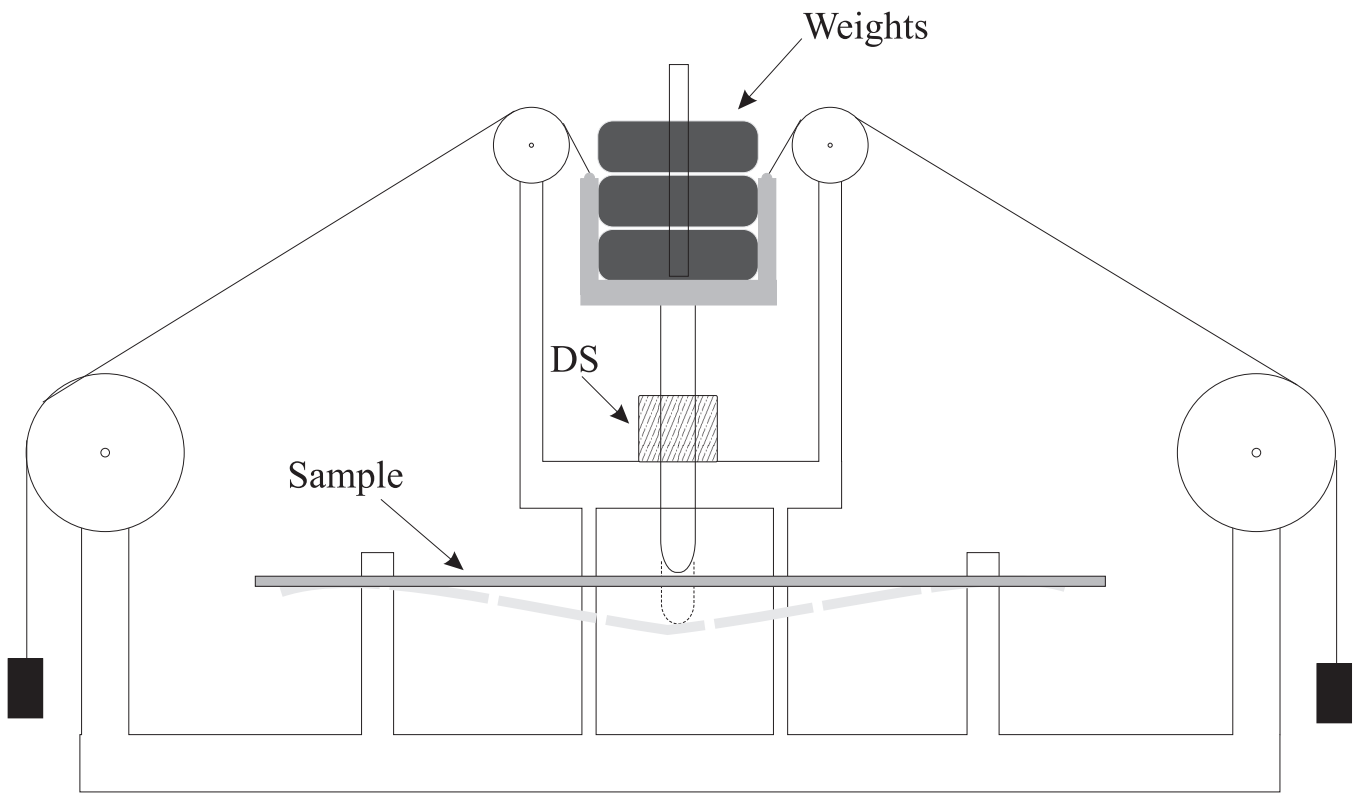}}
\caption{The bending machine (BM). A vertical force is imposed in
the center of the sample by using weights (up to $65\ Kg$), so
that the sample is broken by bending. The edges of the sample can
be clamped or free. The displacement is measured by the sensor
DS.} \label{fig:flexmach}
\end{figure}

\begin{figure}[p]
\centerline{\epsfysize=1.1\linewidth \epsffile{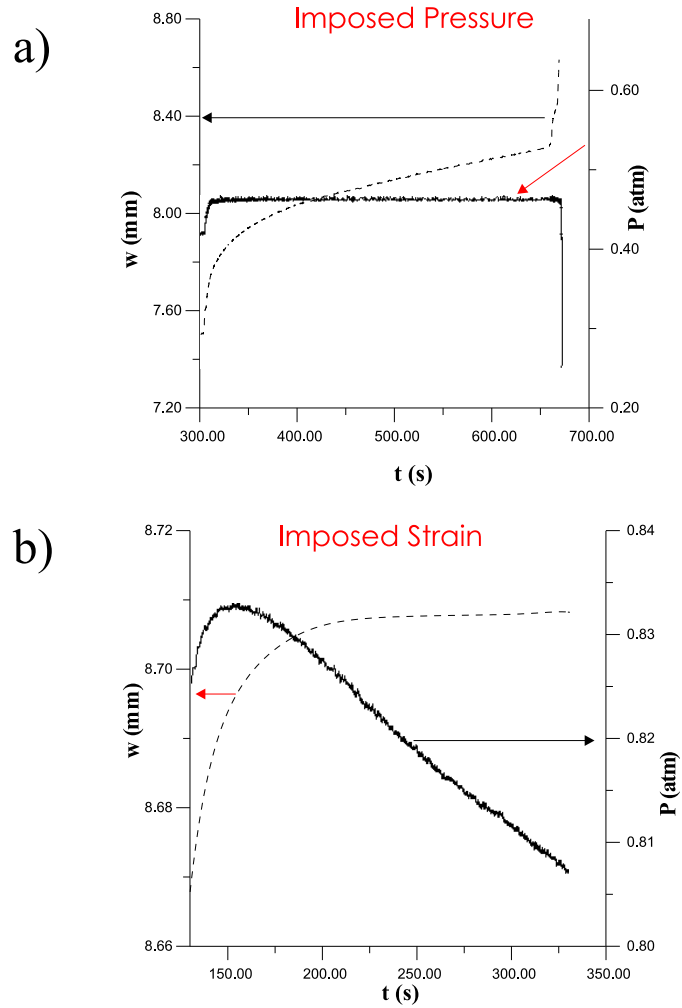}}
\caption{A sample is submitted to a constant load with the HPC. During the load
we measure the pressure (continuous line) and the deformation of the sample in
its center (dashed line). (a) A constant pressure is imposed to the sample.
The deformation increases continuously - even after that the
pressure has reached a constant value - till the sample fails.
(b) A constant deformation is imposed to the sample. In this case, after a
transient period, the pressure decreases till the system reaches a stationary
state (not shown in this picture).} \label{fig:ultima}
\end{figure}

\begin{figure}[p]
\psfig {figure=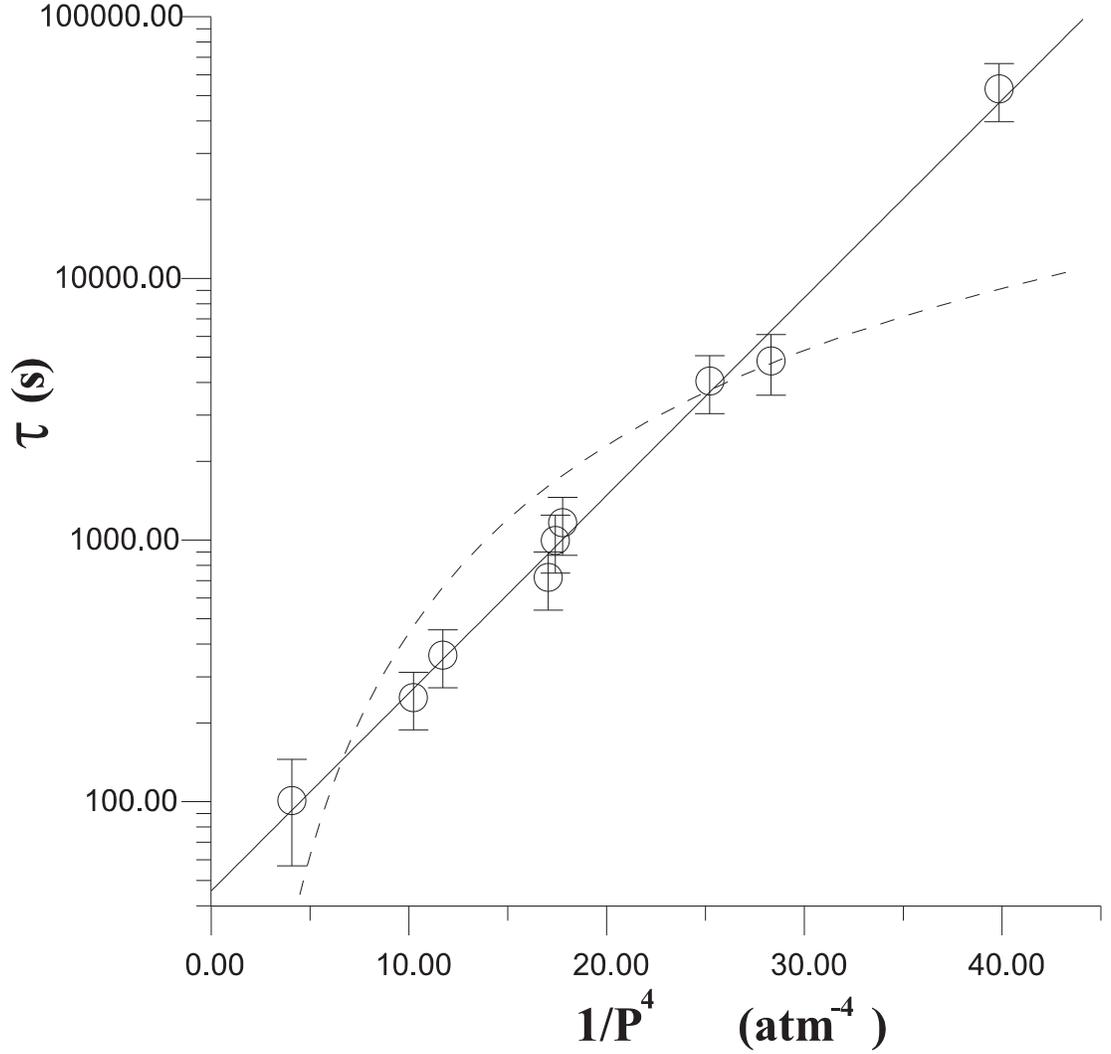, height=14cm} \caption{Measurements on
wood samples. The time $\protect\tau $ needed to break the wood
samples under an imposed constant pressure P is here plotted as a
function of 1/P$^4$ in a semilog scale. The dashed line
represents the solution proposed by Mogi \protect\cite{mogi}($\protect\tau %
=ae^{-bP}$). The continuous line is the solution proposed by Pomeau for
microcrystals ($\protect\tau =\protect\tau _oe^{(P_o/P)^4}$). In the plot $%
\protect\tau _o=50.5$ s and $P_o=0.63$ atm. Every point is the
average of 10 samples. The error bar is the statistical
uncertainty. For the fiberglass samples, we find $\protect\tau
_o=44.6$ s and $P_o=2.91$ atm.} \label{fig:mogi}
\label{fig:energia_cumulata}
\end{figure}

\begin{figure}[tbp]
\centerline{\epsfysize=0.7\linewidth \epsffile{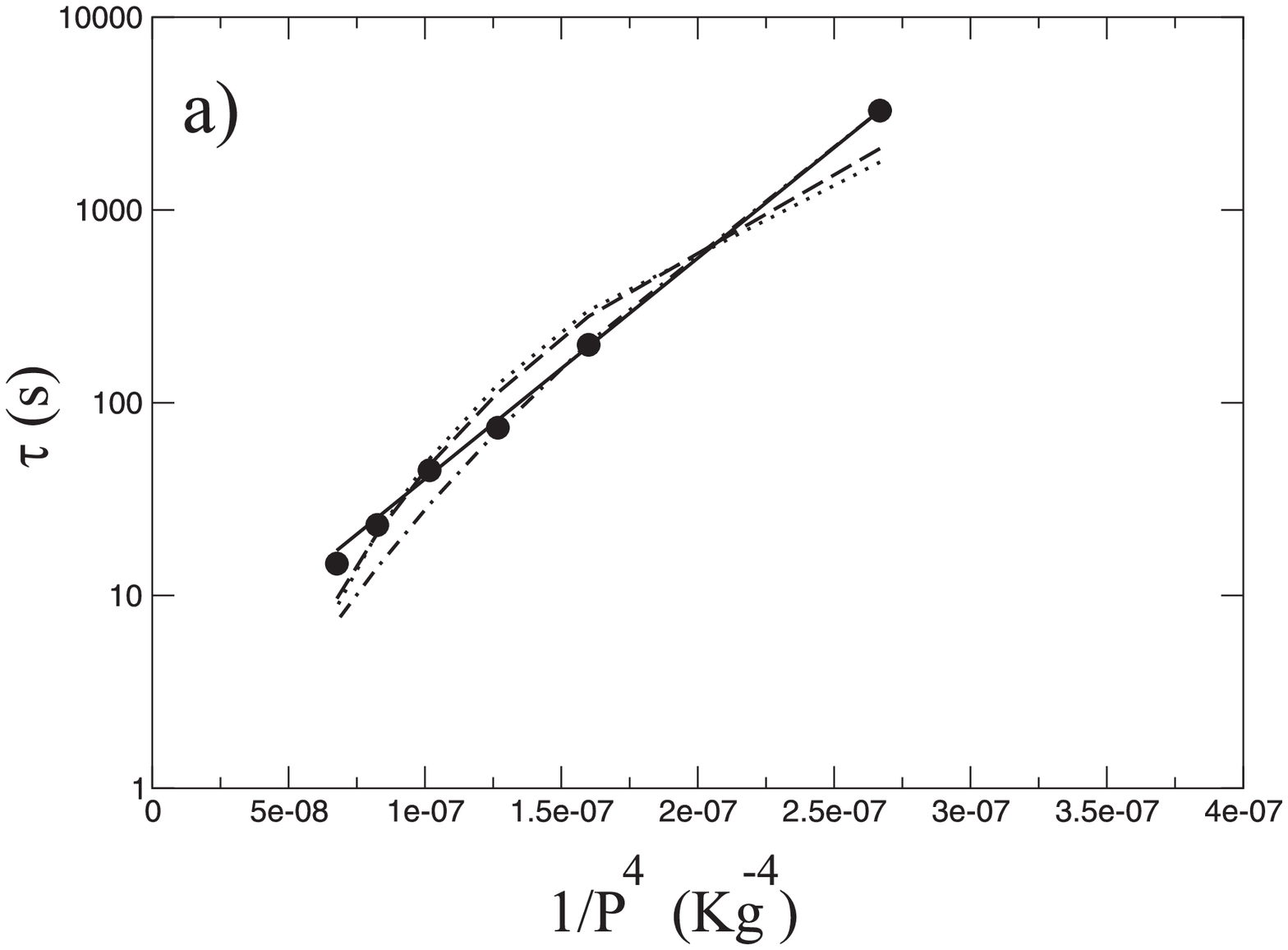}}
\centerline{\epsfysize=0.7\linewidth \epsffile{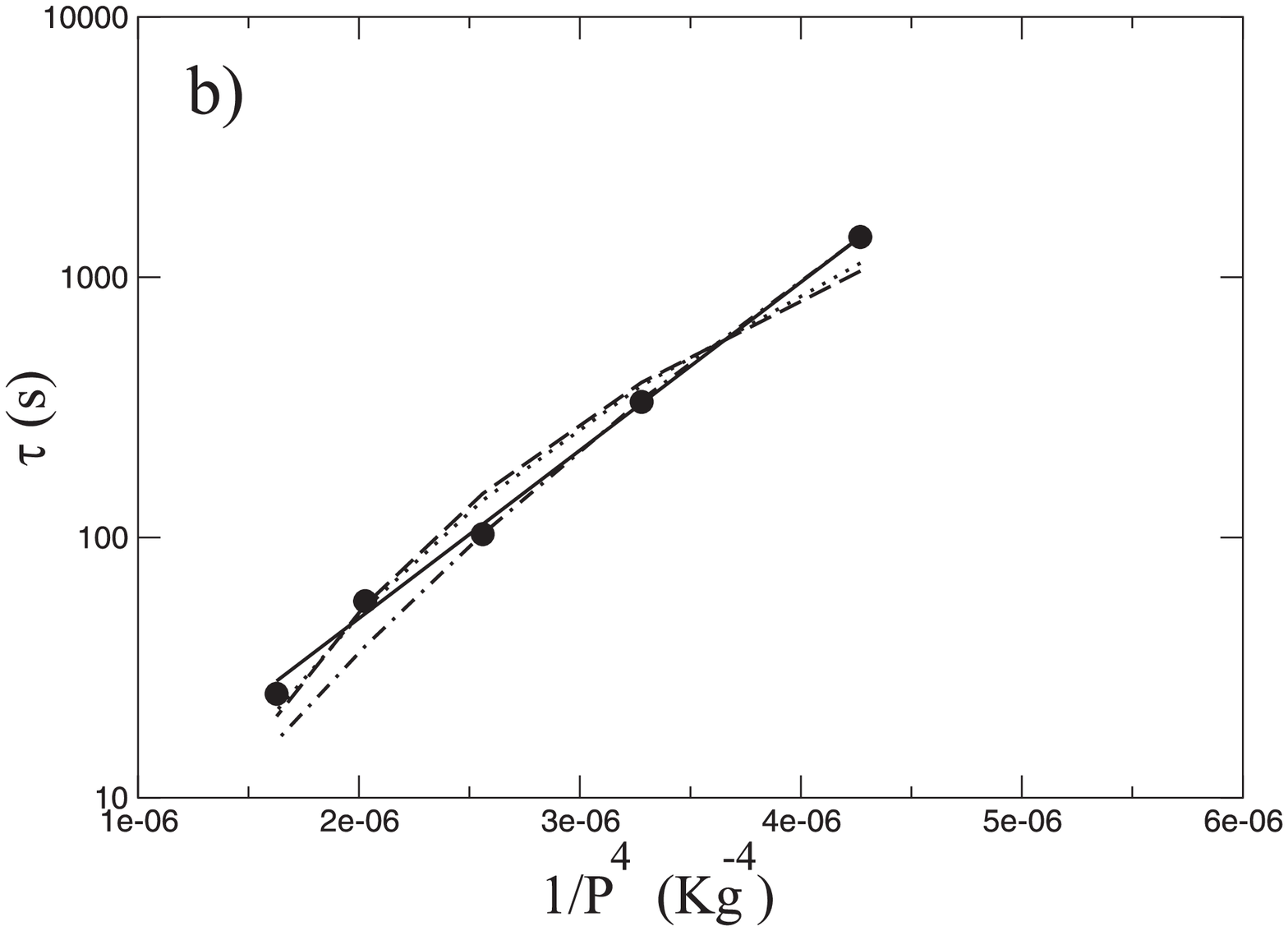}}
\caption{Failure time $\tau$ of the samples in fiberglass broken
with the FM (clamped edges). The sample's size are $22\ \times\ 2\
\times\ 0.2\ cm$ (a) and $22\ \times\ 1\ \times\ 0.2\ cm$ (b).
Each point represents the mean value of 20 measures. Lines
represent the best fit with
$\tau=\tau_o\exp\left(\frac{P_o}{P}\right)^4$ (solid line),
$\tau=\tau_o\exp\left(\frac{P_o}{P}\right)^2$ (dashed-dotted
line), $\tau=A\exp(-bP)$ (dotted line), and $\tau=A\ P^{-b}$
(dashed line).} \label{fig:mariap4a} \label{fig:mariap4b}
\end{figure}

\begin{figure}[tbp]
\centerline{\epsfysize=1.2\linewidth
\epsffile{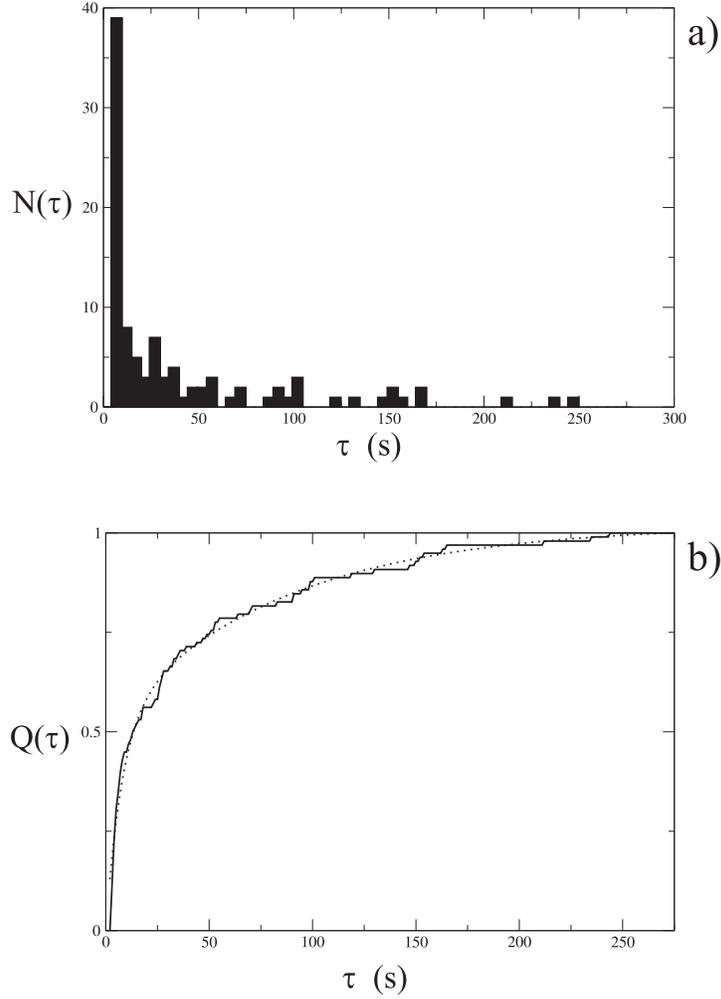}} \caption{Distribution of the
lifetimes $\tau$ of 100 samples in fiberglass broken with the BM
(clamped edges) with a load $P = 54\ Kg $. The sample's size are
$22\ \times\ 2\ \times\ 0.2\ cm$. a) The histogram of $\tau$ shows
that the distribution of lifetimes is not gaussian. b) The
cumulative distribution $Q(\tau) = \frac{\int_0^\tau
N(t)dt}{\int_0^\infty N(t)dt}$ (solid line) is best fitted by the
sum of two exponential terms (dotted line).} \label{fig:mariahist}
\label{fig:mariacumhist}
\end{figure}

\begin{figure}[p]
\psfig {figure=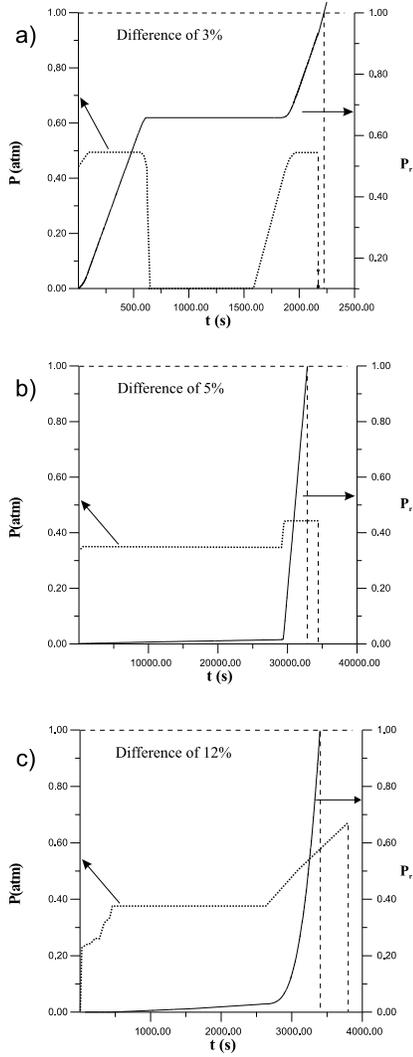, height=14cm} \caption{The imposed
time dependent pressure (bold dotted line) is plotted as a
function of time in the case of wood samples. The continuous line
is
the integral in time of the function $f\left( P\right) ={\frac 1{\protect%
\tau _o}}e^{-P_o^4/P^4}$. On the basis of eq. \ref{integ} the
predicted breaking time $\protect\tau $ is obtained when the
integral of $f\left( P\right) $ is equal to 1. The horizontal
distance between the two vertical dashed lines in each plot
represent the difference between the predicted and the measured
breaking time. In \textbf{a)} a constant pressure has been applied
during about 700 s, then the load is suppressed and then the same
constant load is applied again. The difference between the
life-time predicted by (eq. \ref{integ}) and the experimental
result is of $3\%$. \textbf{b) } Here two pressure plateaux of
different value are successively applied to the sample. The
difference between the measured and the predicted life-time is of
$5\%$. In \textbf{c)} an erratic pressure is applied to the
sample. Here the error is of $10\%$.} \label{fig:tempi_di_rottura}
\end{figure}

\begin{figure}[p]
\psfig {figure=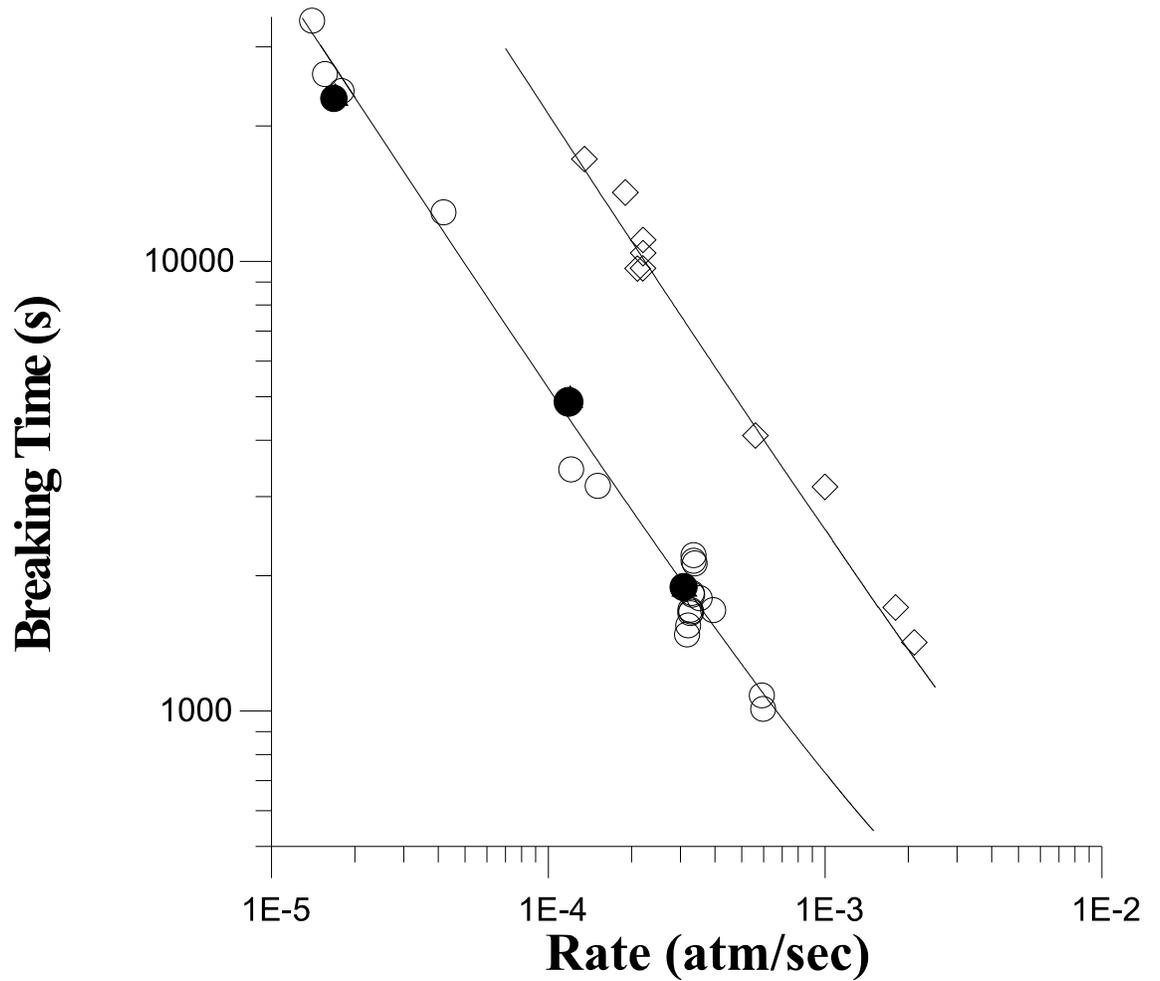, height=13cm} \caption{A load linearly
increasing at different rates $A_p$ has been applied to different
samples. The measured breaking times are plotted as a function of
$A_p$ in a loglog scale; circles and squares represent the
measures on wood and fiberglass samples respectively at T=300 $K$.
Bold circles represent measures on wood samples at T=380 $K$. The
lines are the life time calculated from eq.\ref{integ} using the
best fit values for $P_o$ and $\protect\tau _o$. These experiments
show that eq. \ref{integ} describes well the life-time of the
samples submitted to a time dependent pressure.}
\label{fig:tempi_legno_fibra}
\end{figure}

\begin{figure}[tbp]
\centerline{\epsfysize=1.1\linewidth \epsffile{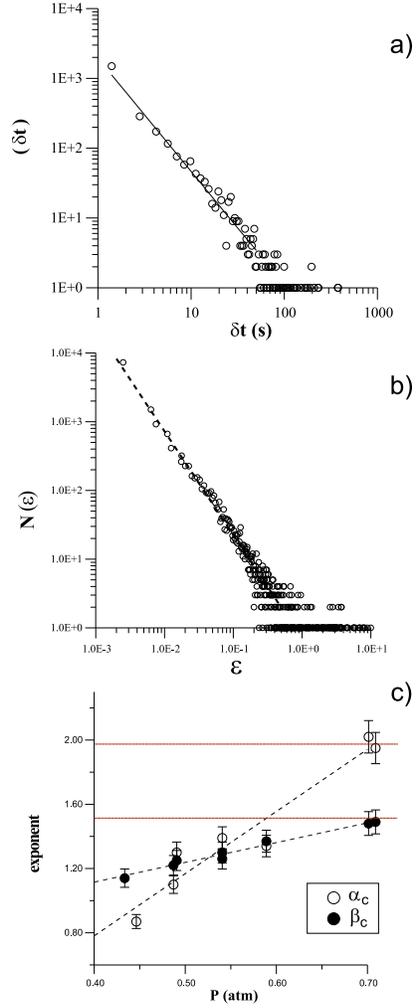}}
\caption{\textbf{a,b)} Two typical time $\protect\delta t$ and energy $%
\protect\varepsilon $ distributions obtained at imposed constant pressure ($%
P=0.56$ $atm$). \textbf{c)} The exponents $\protect\alpha _c$ (empty
circles) and $\protect\beta _c$ (black points), plotted as a function of the
value of the imposed constant pressure. Note that as the pressure increases,
the values of the exponents tend to those obtained in the case of constant
pressure rate. The error bars represent the statistical uncertainty.}
\label{fig:hist}
\end{figure}

\begin{figure}[tbp]
\centerline{\epsfysize=0.9\linewidth \epsffile{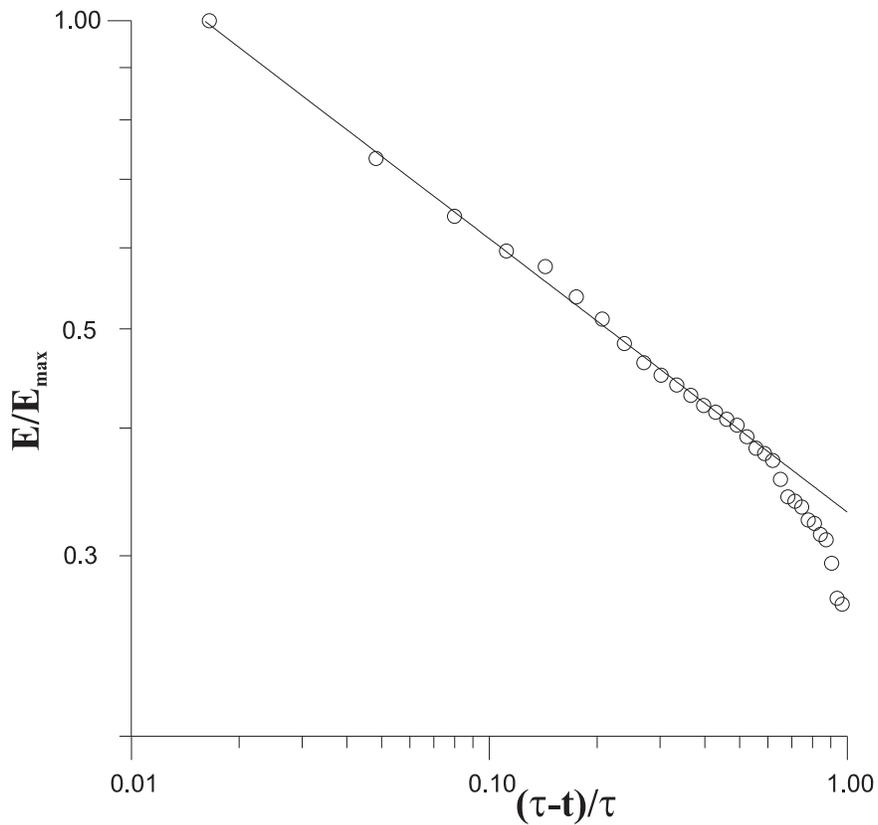}}
\caption{ The cumulative energy E, normalized to E$_{\max }$, as a
function of the reduced control parameter $\frac{\protect\tau
-t}\protect\tau $ at the neighborhood of the fracture point (Case
of imposed constant
pressure). The circles are the average for 9 wood samples.
The solid line is the fit $%
E=E_0\left( \frac{\protect\tau -t}{\protect\tau} \right)
^{-\protect\gamma }$. The exponent found, $\protect\gamma =0.26$,
does not depend on the value of the imposed pressure. In the case
of a constant pressure rate the same law has been found \protect\cite{articolo,prl}.}
\label{fig:critic2}
\end{figure}

\begin{figure}[tbp]
\centerline{\epsfysize=0.9\linewidth \epsffile{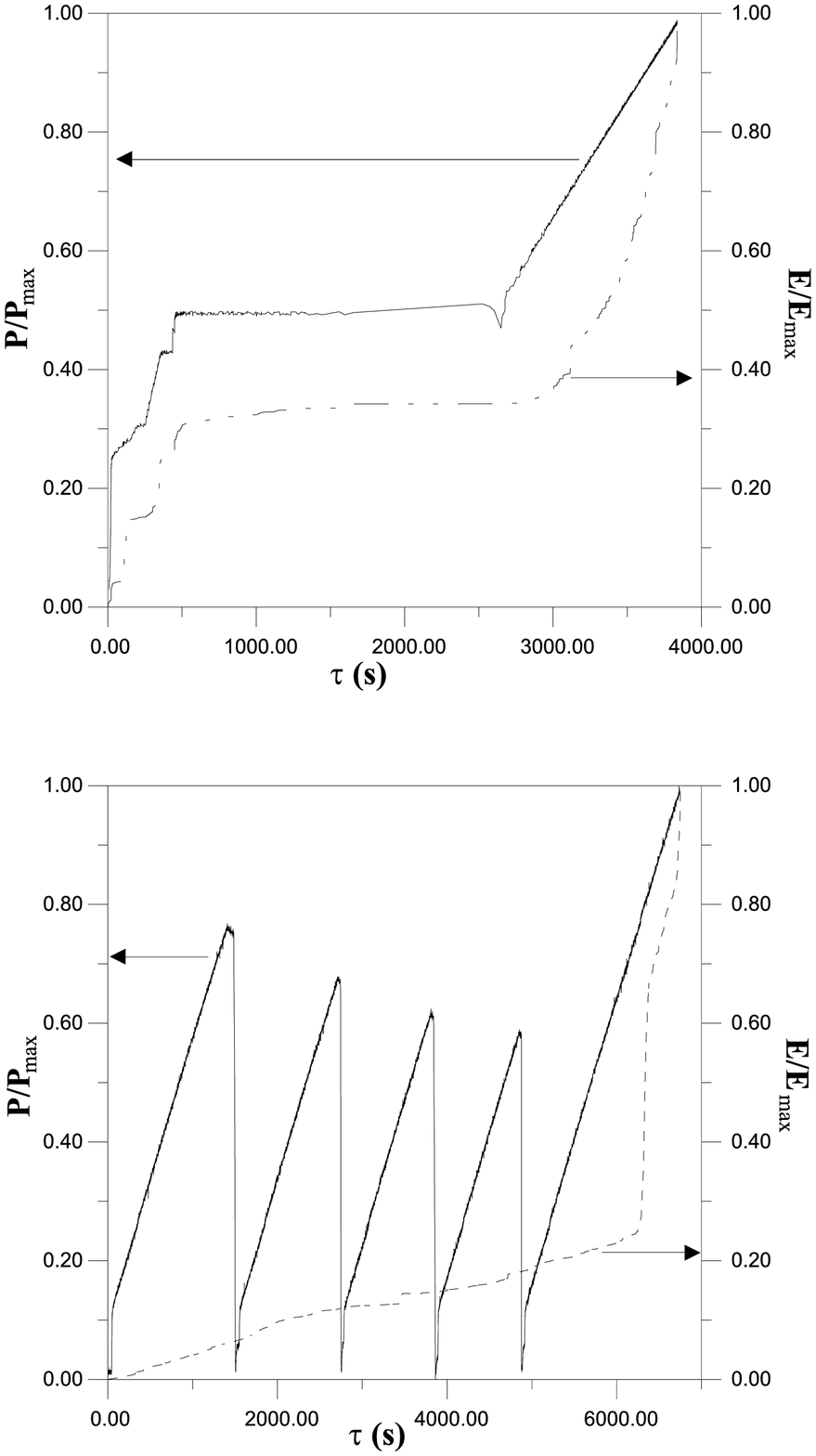}}
\caption{ The imposed pressure, normalized at P$_{\max }$,(solid
line) and the cumulative
energy $E$, normalized to E$_{\max }$, (dashed line) are plotted as a function of time $t$%.
a) An example of erratic pressure. b) A cyclic pressure.}
\label{fig:press}
\end{figure}

\begin{figure}[tbp]
\centerline{\epsfysize=0.9\linewidth \epsffile{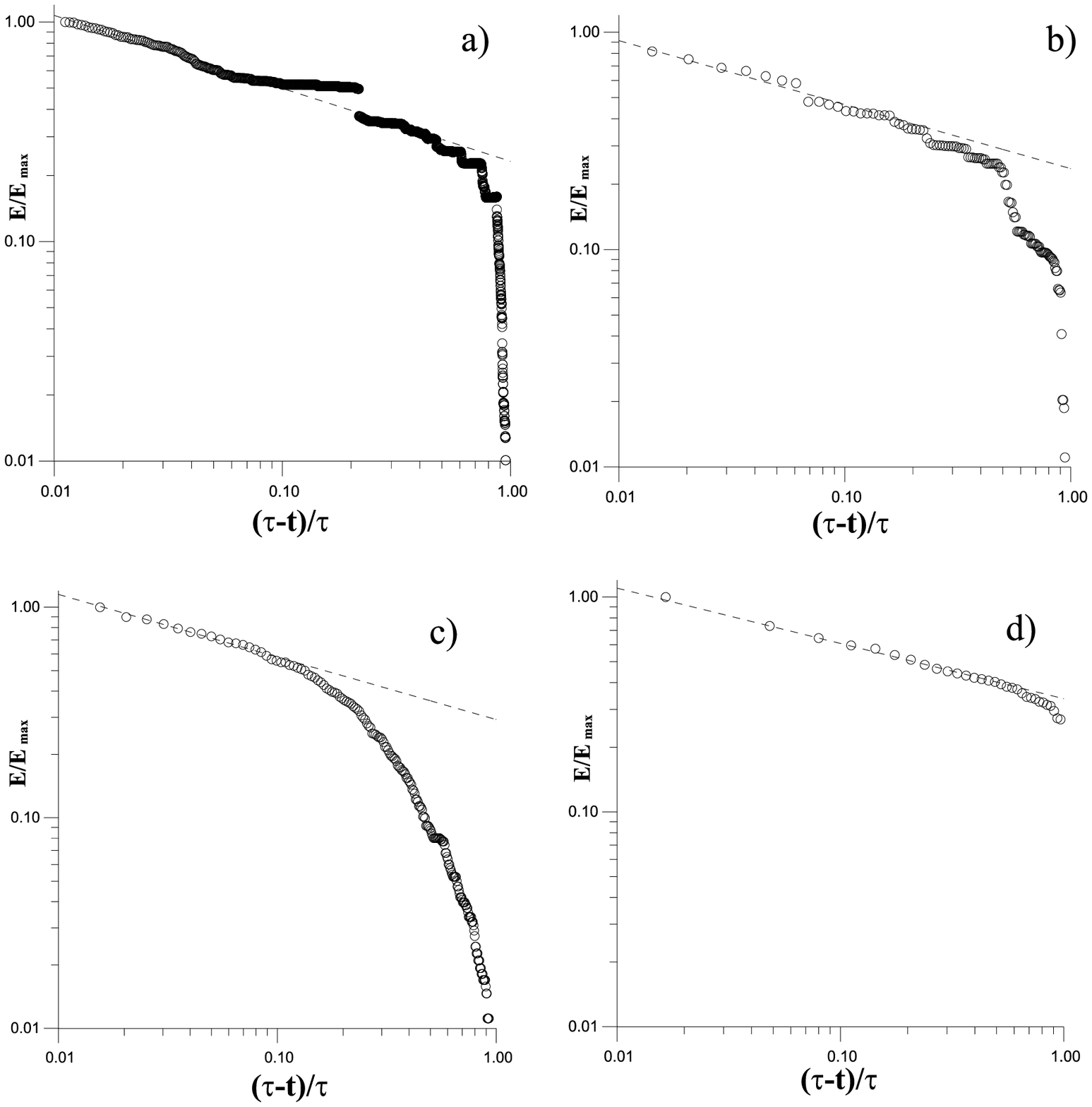}}
\caption{ The cumulative energy $E$, normalized to $E_{\max }$, as
a function of the reduced control parameter $\frac{\protect\tau
-t}{\protect\tau} $ at the neighborhood of the fracture point.
Figure d) represent the measure taken, at imposed constant rate
force, on the tensile machine. The other figures represent
measures made on the HPC apparatus at : imposed constant pressure
(c), imposed cyclic pressure (a) and imposed erratic pressure (b).
The dotted lines are the fit $E=E_0\left( \frac{\protect\tau -t}{\protect\tau} %
\right) ^{-\protect\gamma }$. The exponents found are: $\protect\gamma =0.29$
(a), $\protect\gamma =0.25$ (b), $\protect\gamma =0.29$ (c) and $\protect%
\gamma =0.27$ (d). In the case of a constant pressure rate (on the HPC
machine) the same law has been found \protect\cite{articolo,prl}}
\label{fig:critic}
\end{figure}


\begin{thebibliography}{99}
\bibitem{commentaccio} The thickness of the crystal is just a few
molecular layers.
\bibitem{commento} Several functions with many free parameters have been
proposed in order to fit the dependence of the sample life time on
the applied stress. Among the fitting functins, which are
supported by a microscopic physical mechanism we show in fig.
\ref{fig:mogi}a the two functions which give the best fit of our
data.
\end{thebibliography}
\end{document}